\documentstyle[12pt,epsf]{article}

\begin{document}
\sloppy
\begin{titlepage}
\hfill LMU-WAG-940530 \\
\begin{center}
{\LARGE Fluctuating Interfaces in Microemulsion and Sponge Phases} \\
\vspace{0.5cm}
{\large G. Gompper and J. Goos} \\
{\normalsize \sl 
Sektion Physik der Ludwig-Maximilians-Universit\"at M\"unchen } \\
{\normalsize \sl Theresienstr. 37, 80333 M\"unchen, Germany} \\
{\normalsize \sl May 30, 1994} \\
\end{center}
\vspace{0.5cm}
\begin{abstract}
A simple Ginzburg-Landau theory with a single, scalar order parameter
is used to study the microscopic structure of microemulsions and
sponge phases. The scattering intensity in both film and bulk
contrast, as well as averages of the internal area $S$, the Euler
characteristic $\chi_E$, and the mean curvature squared $<H^2>$, are
calculated by Monte Carlo methods. The results are compared with
results obtained from a variational approach in combination with the
theory of Gaussian random fields and level surfaces. The results for
the location of the transition from the microemulsion to oil/water
coexistence, for the scattering intensity in bulk contrast, and for
the dimensionless ratio $\chi_E V^2/S^3$ (where $V$ is the volume) are
found to be in good quantitative agreement. However, the variational
approach fails to give a peak in the scattering intensity in film
contrast at finite wavevector, a peak which is observed both in the
Monte Carlo simulations and in experiment. Also, the variational
approach fails to produce a transition from the microemulsion to the
lamellar phase. 
\end{abstract}

\begin{center}
\end{center}
\vfill
submitted to Phys. Rev. E
\end{titlepage}

\section{Introduction}

 The understanding of binary and ternary amphiphilic systems has
made considerable progress over the last few years
\cite{books,GSreview}. One of the phases which has received particular
attention is the microemulsion, a homogeneous, isotropic mixture of
oil, water and amphiphile. It is by now well established that the
microemulsion consists of homogeneous regions of oil and water, which
form a complicated, intertwined network, with a typical length scale
of a few hundred \AA. This is possible because the amphiphile forms a
monolayer at the interface between these oil- and water-regions and
thereby reduces the interfacial tension, so that a phase with an
extensive amount of internal interface can become stable. The
structure of the sponge phase in aqueous surfactant solutions is very
similar \cite{Porte,sponge}. In this case an amphiphilic bilayer
separates two multiply connected water networks.

 In order to characterize the internal structure of a
microemulsion, several quantities have been proposed. Most experiments
and theoretical studies have concentrated on the water-water
correlation function, or equivalently the scattering intensity in bulk
contrast \cite{books,GSreview}. However, a two-point correlation
function gives little information about the connectivity and
percolation of the oil- and water-networks in a microemulsion. The
latter property can be probed by conductivity and diffusion
experiments. From a theoretical point of view, the topology of a
microemulsion can be characterized by its Euler characteristic
\cite{HofKlein,HuseLeib,Teubner90,MeckeWag}, a quantity which
unfortunately is difficult to measure experimentally \cite{Teubner90}.

 We want to investigate here a simple Ginzburg-Landau model for
oil-water-amphiphile mixtures \cite{TeuStrey,GoSch90}. This model has
been used previously to calculate the spectrum of capillary waves of
an oil/water interface \cite{GoSch90,GoKrausI}, to explain the wetting
behavior of the microemulsion at the oil/water interface
\cite{GoSch90,GoHoSch,PuHoSch}, to describe several ordered phases
like a lamellar, a hexagonal and a cubic phase \cite{GoZsch}, and to
study the behavior of amphiphilic systems in confined geometry
\cite{GoKrausI,SchmidSch}. The same model has also been used to
predict sound attenuation and dispersion in microemulsions
\cite{GoHennes94}. In Ref. \cite{GoKrausII} this model has been
studied by Monte Carlo simulations. It has been shown that there is a
region in the phase diagram, where the microemulsion is stabilized by
thermal fluctuations. This is the part of the phase diagram we want to
investigate here in more detail. In particular, simulations are used
to calculate the scattering intensity both in bulk and film contrast,
and the geometrical quantities area, Euler characteristic and the mean
curvature squared. The results are then compared with the results
obtained from a simple variational calculation and from a
self-consistent perturbation theory. 

 The variational approach has been applied recently to calculate
the scattering intensity of an ensemble of random interfaces with
bending rigidity \cite{PierSaf}. To make the random interface model
accessible to the variational method, a mean-spherical approximation
has been used as a first step. In comparison to that model, we are
here in the unique position that we can directly compare the results
obtained from the variational approach with the results of our Monte
Carlo simulation, and thus check the quality of this approximation.

\section{Ginzburg-Landau Model}

 Our analysis is based on the free-energy functional
\cite{GoSch90}
\begin{equation}\label{GL} 
{\cal F} \lbrace \Phi \rbrace = \int d^{3}r\ 
\left( c (\Delta \Phi)^{2} + g(\Phi) (\nabla \Phi)^{2} 
                 + f(\Phi) - \mu\Phi \right )  
\end{equation}
for a single, scalar order parameter field $\Phi({\bf r})$, which is
proportional to the local difference of the oil and water
concentrations. Here, $\mu$ is the chemical potential difference
between oil and water. The amphiphile concentration does not appear
explicitly in our model, and should be considered to be integrated out
\cite{CJPW90}. The average amphiphile concentration enters the model
(\ref{GL}) via the parameters of the functions $f$ and $g$, which are
chosen to take the form \cite{GoKrausII}
\begin{eqnarray}\label{f}
f(\Phi) & = & \omega (\Phi - \Phi_{o})^{2}(\Phi^{2} + f_{0})
                                  (\Phi - \Phi_{w})^{2} \\
\label{g}
g(\Phi) & = & g_{0} + g_{2} \Phi^{2} \ .
\end{eqnarray}
With this choice for $f$ and $g$, a three-phase coexistence between a
oil-rich phase with ${<\Phi>} \simeq \Phi_o$, a water-rich phase with
$<\Phi> \simeq \Phi_w$, and a microemulsion with $<\Phi> \simeq 0$ can
be described. We consider here only systems with oil-water symmetry,
where $\Phi_o=-\Phi_w \equiv \Phi_b$. The constant $f_0$ in Eq.
(\ref{f}) is proportional to the chemical potential of the amphiphile,
while $g_0$ decreases with increasing amphiphile strength or
amphiphile concentration \cite{Lerczak}. The value of $g_0$ determines
the behavior of correlation function $G(r) = <\Phi({\bf r})\Phi(0)>$,
and of the scattering intensity in bulk contrast
\cite{TeuStrey,GoSch90}. For $g_0$ sufficiently small or negative, the
correlation function in the microemulsion decays with damped
oscillations, while the scattering intensity has a peak at non-zero
wavevector $k$. To ensure a monotonic decay of $G(r)$ in the
oil-rich and water-rich phases, we choose
\begin{equation}
g_{2} = 4 \sqrt{1 + f_0} - g_{0} + 0.01 \ .
\end{equation}
The natural length scale of the model (\ref{GL}), (\ref{f}), (\ref{g})
is \cite{GoHoSch}
\begin{equation}
\ell_0 = \left( \frac{c}{\omega \Phi_b^4} \right)^{1/4} \ .
\end{equation}
For all explicit calculations, we use the parameter set $c=1$,
$\omega=1$, $\Phi_b=1$, and $\mu=0$; this implies in particular
$\ell_0=1$. Thus, we study the microemulsion structure as a function
of the parameters $f_0$ and $g_0$. 

 The same model can also be interpreted as a model for binary
amphiphilic systems \cite{CRAMS88,RCOBNB,sponge,GoSch94}, if the
amphiphilic molecules form bilayers without holes or seams. The
bilayers separate space into an "inside" ($\Phi>0$) and an "outside"
($\Phi<0$). In this case, the microemulsion phase corresponds to the
symmetric sponge phase, and the oil-rich and water-rich phases to the
asymmetric droplet phase.

\section{Methods}
\subsection{Monte Carlo Simulations}

 To study the field-theoretic model (\ref{GL}) by Monte Carlo
simulations, space has to be discretized by introducing a simple cubic
$N\times N\times N$ lattice with lattice constant $a_0$ (and periodic
boundary conditions). However, the order parameter field $\Phi({\bf
r}_i)$ at each lattice site $i$ is still a continuous, real variable.
Details about the simulation procedure, about the triangulation of the
$\Phi({\bf r})=0$ surface, and about the calculation of the area can
be found in Ref. \cite{GoKrausII}. In addition to the quantities
\cite{Euler} studied in Ref. \cite{GoKrausII}, we also calculate here
the mean curvature squared and the scattering intensity in film
contrast. 

 The calculation of the film scattering is rather straightforward.
As in models for level surfaces and Gaussian random fields, we define
the film scattering intensity as the Fourier transform of the
correlation function
\begin{equation}
G_{film}(r; \epsilon) = 
   N_0 < \delta_\epsilon(\Phi({\bf r})) \ \delta_\epsilon(\Phi(0)) >
\end{equation}
where $\delta_\epsilon(x) = 1/\epsilon$ for $-\epsilon/2 \le x \le
+\epsilon/2$, and zero otherwise. The normalization factor $N_0$ is
chosen such that $\lim_{r\to\infty} G_{film}(r; \epsilon) = 1$. This
correlation function is proportional to the probability distribution
to find two points at distance $r$ in a thin layer of thickness
$\epsilon$ of the $\Phi=0$ surface. In the simulations, we  calculate
the distance distribution of lattice sites for which $|\Phi({\bf r})|
\le \epsilon/2$ (with periodic boundary conditions).

 To calculate the mean curvature, we consider two neighboring
triangles $j$ and $k$. Let $b_{jk}$ be their common edge, $U_j$ and
$U_k$ their perimeters, and $S_j$ and $S_k$ theirs areas. Each
triangle has a normal vector ${\bf n}$ pointing towards the oil.  For
the calculation of the radius of curvature, we have to assume that the
two triangles are approximately equilateral. The radius of the sphere,
which touches both triangles in their centers of mass, is then taken
to be the radius of curvature, 
\begin{equation} \label{R_jk}
|R_{jk}| =  3^{-3/4} \left( \frac{S_j+S_k}{2} \right)^{1/2} 
\left( \frac{ 2 - {\bf n}_j \cdot {\bf n}_k } 
              { {\bf n}_j \cdot {\bf n}_k } \right)^{1/2} \ .
\end{equation}
The sign of $R_{jk}$ is easily obtained from the two normal vectors,
and the vector connecting the centers of mass of the two triangles.
The mean curvature $c_j$ of the whole triangle $j$ is obtained by a
sum over the contributions from the edges with the three neighboring
triangles $k_1$, $k_2$, and $k_3$, weighted with the lengths of these
edges,
\begin{equation} \label{c_j}
c_j =  \frac{1}{U_j} \sum_{i=1}^{3} \frac{b_{jk_i}}{R_{jk_i}} \ .
\end{equation}
We want to point out that it is essential to define the mean curvature
on triangles rather than on edges, because only in this way can the
mean curvature near saddle points be described correctly. 

 The total mean curvature and the total mean curvature squared are 
then sums over all triangles,
\begin{eqnarray} \label{curv}
\int dS \ H &=& \frac{2}{\sqrt{3}} \sum_{j} S_j c_j \\
\label{curv_square}
\int dS \ H^2 &=& \frac{2}{\sqrt{3}} \sum_{j} S_j c_j^2 \ . 
\end{eqnarray}
where the integral is over the whole surface. The additional factor
$2/\sqrt{3}$ is obtained by triangulating the surface of a sphere and
comparing the results (\ref{curv}) and (\ref{curv_square}) with the
exact results in the limit of large sphere radius. A similar procedure
has been used in Refs. \cite{SeungNels} and \cite{KroGo92}.

 In order to test the quality of these expressions, we have
calculated the (average) mean curvature and mean curvature squared for
spheres and cylinders, as a function of the radius $R$. For $R>2a_0$,
where $a_0$ is the lattice constant, the agreement with the exact
results is very good, with deviations of only a few percent. We have
also studied corrugated surfaces (with saddle points), which are given
by the Monge parametrization $z(x,y) = z_0 \sin(2\pi x/L) \sin(2\pi
y/L)$. In this case, the deviations are less than a few percent for 
$L>6a_0$ with $z_0=a_0$, and for $L>9a_0$ with $z_0=3 a_0$.

 An important question in any Monte Carlo simulation is if the
system has reached thermal equilibrium. To answer this question, we
have calculated the auto-correlation function \cite{MC} in thermal
equilibrium, as a function of Monte Carlo time $t$, 
\begin{equation} \label{auto}
\tilde G(t) = \frac{\sum_{i=1}^{N^3} 
\left[ <\Phi({\bf r}_i, t) \Phi({\bf r}_i, 0) > - <\Phi(0, 0)>^2 \right] }
{ N^3 \left[ <\Phi(0, 0)^2> - <\Phi(0, 0)>^2 \right] }
\end{equation}
where $N^3$ is the number of lattice sites. The exponential decay of
this function gives us the relaxation time $\tau$. The result for
$\tilde G(t)$ for our model in the microemulsion, near the point of
four-phase coexistence of oil-rich, water-rich, microemulsion and
lamellar phases, is shown in Fig. 1. It shows that the relaxation time
is $\tau \simeq 680$ (in units of Monte Carlo steps per lattice site
(MCS)). Since in our simulations we usually average over $50000$ to
$100000$ MCS, all our results represent true equilibrium averages.

\subsection{Variational Approach}

 The variational method \cite{variation} introduces a
Gaussian model with the free energy functional
\begin{equation} \label{Gauss}
{\cal F}_0\{\Phi\} = \int d^3r \int d^3r' \ 
     (\Phi({\bf r})-\overline\Phi) \ G_0^{-1}(|{\bf r-r'}|) \ 
                       (\Phi({\bf r}')-\overline\Phi) \ .
\end{equation}
Then the Feynman-Bogoliubov inequality
\begin{equation} \label{Bogoliubov}
F \le F_u \equiv F_0 + <{\cal F}-{\cal F}_0>_0
\end{equation}
is used to construct an upper bound for the  free energy $F$ of the
system described by the functional ${\cal F}$. Here, both the free
energy $F_0$ and the average $<...>_0$ are calculated with the free
energy functional (\ref{Gauss}). In Eqs. (\ref{Gauss}) and
(\ref{Bogoliubov}), a cutoff $\Lambda$ in momentum space is implicit.
In order to compare with the results of the Monte Carlo simulations,
we set $\Lambda = 2\pi/(\gamma_0 a_0)$, where $a_0$ is the lattice
constant used in the simulations, and $\gamma_0 \simeq 1$. The best
approximation to $F$ in this approach is then obtained by minimizing
$F_u$ with respect to $G_0$ and $\overline\Phi$.  For the free energy
functional (\ref{GL}) with (\ref{f}) and (\ref{g}), we find 
\begin{eqnarray}
F_u &=& 15 \omega G_0(r=0)^3 + A_2 G_0(r=0)^2 + A_1 G_0(r=0) + A_0 \\
    &+& \int \frac{d^3 k}{(2\pi)^3} 
      \left\{ G_0(k) \left[c k^4 + (B + g_2 G_0(r=0) k^2) \right] 
            - \frac{1}{2} \ln(G_0(k)) \right\}  \ , \nonumber
\end{eqnarray}
where
\begin{eqnarray}
A_2 &=& 3\omega (15 {\overline \Phi}^2 + f_0 - 2\Phi_b^2) \\
A_1 &=& \omega \left[ 15 {\overline \Phi}^4 
         + 6(f_0-2\Phi_b^2) {\overline \Phi}^2 
         + (\Phi_b^2 - 2f_0)\Phi_b^2 \right] \\
A_0 &=& \omega ({\overline \Phi}^2 + f_0)({\overline \Phi}^2 
                                  - \Phi_b^2)^2 \\
B &=& g_0 + g_2{\overline \Phi}^2 \ .
\end{eqnarray}
The optimal Gaussian correlation function is obtained by requiring
that the functional derivative of $F_u$ with respect to $G_0(k)$ must
vanish, which implies
\begin{equation} \label{EL_var}
[45 \omega G_0(r=0)^2 + 2 A_2 G_0(r=0) + A_1] + [B + g_2 <k^2>] k^2
    + c k^4 - \frac{1}{2G_0(k)} = 0 \ ,
\end{equation}
where $<k^2>$ is the second moment of $G_0(k)$ in Fourier space. Thus, 
the correlation function in the variational approach has the
Teubner-Strey form \cite{TeuStrey}
\begin{equation} \label{corr_var}
\ G_0(k)^{-1} = 2(b_0 + b_2 k^2 + c k^4) \ ,
\end{equation}
with constants $b_0$ and $b_2$. Its Fourier transform is given by
\cite{TeuStrey}
\begin{equation} \label{corr_posit}
G_0(r) = A \frac{1}{r} e^{-r/\xi} \sin{qr}
\end{equation}
where $\xi$ is the correlation length, with
\begin{equation} \label{xi}
\xi^{-2} = \frac{1}{2} \sqrt{\frac{b_0}{c}} + 
                           \frac{1}{4} \frac{b_2}{c} \ ,
\end{equation}
$2\pi/q$ is the average domain size of coherent oil- and
water-regions, with
\begin{equation} \label{domain}
q^{2} = \frac{1}{2} \sqrt{\frac{b_0}{c}} - 
                        \frac{1}{4}\frac{b_2}{c} \ , 
\end{equation}
and 
\begin{equation} \label{amplitude}
A = \frac{\xi}{16\pi c q} \ .
\end{equation}
The correlation function (\ref{corr_var}) in combination with the
theory of Gaussian random fields \cite{Berk,Teubner91,PierMarc} can
then be used to calculate the film scattering intensity, and the film
area (per unit volume) $S/V$, the mean curvature squared $<H^2>$, and
the Euler characteristic (per unit area) $\chi_E/S$. For a film of
thickness $\epsilon$, the film scattering intensity is found to be
\cite{PierMarc}
\begin{eqnarray} \label{film_scatt_eps}
G_{film}^{(0)}(r;\epsilon) &=& 
     \frac{N_0}{2\pi <\Phi^2>} \ \frac{1}{\sqrt{1-g(r)^2} } \\
 & & \int_{-\infty}^{+\infty} ds \int_{-\infty}^{+\infty} dt \ 
      \exp\left[ -\frac{s^2+t^2-2g(r)st}{2<\Phi^2>(1-g(r)^2)} \right]
                    \delta_\epsilon(s) \delta_\epsilon(t) \nonumber
\end{eqnarray}
where $g(r) \equiv G_0(r)/G_0(r=0)$.
In the limit $\epsilon \to 0$, this expression simplifies to
\begin{equation} \label{film_scatt_0}
G_{film}^{(0)}(r) = \frac{1}{\sqrt{1-g(r)^2}} \ ,
\end{equation}
or explicitly, with (\ref{corr_posit}),
\begin{equation} \label{film_posit}
G_{film}^{(0)}(r) = 
\left[1-\exp(-2r/\xi) \frac{\sin(qr)^2}{(qr)^2} \right]^{-1/2} \ .
\end{equation}
Thus, the film correlation function diverges as $G_{film}^{(0)}(r)
\sim \sqrt{\xi/r}$ for $r\to 0$. Note that it follows immediately from
Eq. (\ref{film_scatt_0}) that $G_{film}^{(0)}(r) \ge
G_{film}^{(0)}(r=\infty)$. 

 To obtain the geometrical quantities, we employ the exact results
\cite{Teubner91} for $\Phi({\bf r}) = {\overline \alpha}$ surfaces of
Gaussian random fields with $<\Phi> = 0$ \cite{general_barphi},
\begin{eqnarray} \label{random_S} 
S/V &=& \frac{2}{\sqrt{3}\pi} \sqrt{<k^2>} \
              \exp \left( -\frac{\alpha^2}{2} \right) \\
\label{random_K}
\chi_E/S &=& \frac{1}{12\pi} <k^2> (\alpha^2-1) \\
\label{random_H}
<H> &=& \frac{\sqrt{\pi}}{2\sqrt{6}} \sqrt{<k^2>} \alpha \\
\label{random_H2}
<H^2> &=& \frac{1}{6} <k^2> 
      \left( \frac{6}{5} \frac{<k^4>}{<k^2>^2} + \alpha^2 - 1 \right)
\end{eqnarray} 
where $\alpha = {\overline\alpha}/\sqrt{G_0(r=0)}$.
Here, $<k^2>$ and $<k^4>$ are the second and fourth moments of the 
correlation function $G_0(k)$ in Fourier space.

\subsection{Self-Consistent Perturbation Theory}

 Another method to calculate correlation functions beyond the
Ornstein Zernike (OZ) approximation is a perturbation in the
higher-than-quadratic terms in (\ref{GL}) with (\ref{f}) and
(\ref{g}). The two-point correlation function can be expressed by the
Dyson equation,
\begin{equation} \label{dyson}
G(k)^{-1} = G_{OZ}(k)^{-1} - \Sigma(k)
\end{equation}
in terms of the self-energy $\Sigma(k)$. Here,
\begin{equation}
G_{OZ}(k)^{-1} = 2( ck^4 + g_0 k^2 + \omega)
\end{equation}
is the bare propagator. The self-energy $\Sigma(k)$ can now be
expressed in terms of the full propagator $G(k)$, and the full
four-point and six-point vertex functions $\Gamma^{(4)}(k)$ and
$\Gamma^{(6)}(k)$. The Feynman diagrams are shown in Fig. 2a. To
proceed, we need an expression for the vertex functions. This can only
be done approximately. The simplest approximation is to use the bare
vertex functions (see Eqs. (\ref{GL}), (\ref{f}), (\ref{g}))
\begin{eqnarray}
\Gamma_{bare}^{(4)}(k_1,k_2,k_3,k_4) &=& 
    24 \left[ \frac{g_2}{12} (k_1^2 + k_2^2 + k_3^2 + k_4^2) 
     + \omega(f_0-2\Phi_b^2) \right] \\
   & & {\hskip 4 true cm}  \delta(k_1 + k_2 + k_3 + k_4) \nonumber \\
\Gamma_{bare}^{(6)}(k_1,k_2,k_3,k_4,k_5,k_6) &=& 
    720 \omega \ \delta(k_1 + k_2 + k_3 + k_4 + k_5 + k_6)  \ .
\end{eqnarray}
The Dyson equation (\ref{dyson}) is then solved self-consistently.
This is the approximation used in Ref. \cite{LevMunDaw} to calculate
the scattering intensity for a Ginzburg-Landau model (\ref{GL}) with
$g_2=0$ in Eq. (\ref{g}) and $f(\Phi) = t \Phi^2 + v \Phi^4$. In our
case, it gives very bad results in the region of the phase diagram,
where strongly structured microemulsions are stable (neither the peak
position nor the peak intensity are consistent with the Monte Carlo
data). The simple approximation can be improved by calculating the
vertex functions self-consistently. We include all Feynman diagrams up
to two-loop order, as shown in Fig. 2b and 2c, where again all
internal lines correspond to the full propagator $G(k)$. In addition,
we expand the vertex functions in powers of $k$, and truncate this
series after the first few terms, such that the form of the free
energy functional (\ref{GL}), (\ref{f}), (\ref{g}) is recovered after
each step. The resulting equations are solved numerically.

\section{Gaussian Random Fields}

 In order to test our numerical procedures, we have simulated a
model for Gaussian random surfaces, with $g_2=0$ in Eq. (\ref{g}), and
\begin{equation}
f(\Phi) = \omega \Phi^2  \ .
\end{equation}
In this case, the exact results for level surfaces in the continuum
model are given by Eqs. (\ref{random_S}), (\ref{random_K}),
(\ref{random_H}), and (\ref{random_H2}). Our numerical results
are shown in Fig. 3. In all four cases the linear or quadratic
dependence on $\alpha$ is reproduced very well. We have also
calculated the prefactors from Eqs. (\ref{random_S}), (\ref{random_K}),
and (\ref{random_H}). Note that $<k^2>$ has a strong cutoff
dependence, and $<k^4>$ even more so. For the cutoff parameter
$\gamma_0=2.5$, we obtain the full lines in Fig. 3. In the case of
$<H>$, the agreement is excellent. For $S/V$ and $\chi_E/S$, there is
an appreciable deviation for large values of $\overline\alpha$. This
is not very surprising, since the structures get very small in this
case, and thus cannot be described very well by our triangulation
procedure. For $<H^2>$, the amplitude of $\alpha^2$ in Fig. 3c
is about a factor two too large compared to Eq. (\ref{random_H2}).

 Eqs. (\ref{random_S}), (\ref{random_K}), and (\ref{random_H}) can
be used to obtain a relation between $S/V$, $\chi_E$ and $<H>$, which
is {\it independent} of $\alpha$ and $<k^2>$, and thus does not depend
on any cutoff. One easily finds that
\begin{equation} \label{scale_KHS}
- \frac{\chi_E V^2}{S^{3}} = \frac{\pi}{16} \ 
       \Theta\left( - \frac{<H>^2 S}{2\pi\chi_E} \right)
\end{equation}
with $\Theta(0)=1$, where
the scaling function $\Theta$ is given by
\begin{equation} \label{scale_fun}
\Theta(x) = \frac{\pi} {\pi+4x} \ 
                  \exp\left( \frac{4x}{\pi+4x} \right) \ .
\end{equation}
A comparison of the Monte Carlo data for the Gaussian model with the
scaling form (\ref{scale_KHS}) is shown in Fig. 4. The agreement in
this case is very good, with deviations of only a few percent.

\section{Phase Diagram}

 The phase diagram of our Ginzburg-Landau model, calculated in the
mean-field approximation, by Monte Carlo simulations
\cite{GoKrausII}, and with the variational method, is shown in Fig.
5. The results of the variational approach depend on the choice of the
cutoff $\Lambda= 2\pi/(\gamma_0 a_0)$. We have determined $\gamma_0$
from a fit of the scattering intensity to the Monte Carlo data at {\it
one} point in the phase diagram (see discussion below), which gives
$\gamma_0 = 1.5$. It can be seen in Fig. 5 that with this value of
$\gamma_0$, the variational result for the location of the line of
phase transitions from oil/water coexistence to the microemulsion or
lamellar phase reproduces the qualitative behavior the Monte Carlo
data very well; the quantitative agreement is not perfect, but much
better than the mean-field result.

 The lamellar phase cannot be studied with the variational ansatz
(\ref{Gauss}) with an isotropic correlation function $G_0$. Thus, to
detect a transition from the microemulsion to a spatially ordered
phase, we are looking for a spinodal, where the scattering intensity
$G_0(k)$ diverges at some non-zero value of the wavevector $k$. Much
to our surprise, no such spinodal exists. A closer look at the
structure of Eq. (\ref{EL_var}) reveals that this is not a numerical
problem. Since $G_0(r=0) = \xi/(16\pi c)$ (see Eqs. (\ref{corr_posit})
and (\ref{amplitude})), the behavior of $G_0(k)$ for large $\xi$ (where
$<k^2> \simeq q^2$) is given by
\begin{equation}
G_0(k) \simeq \frac{1}{2( c k^4 + (g_0 + r_2) k^2 + r_0 \xi^2 )} \ ,
\end{equation}
with constants $r_0$ and $r_2$. From a calculation similar to that
leading to Eq. (\ref{xi}), one finds that the spinodal is located at
$g_0 + r_2 = -\sqrt{4c r_0} \xi$. Thus, $\xi \to \infty$ implies
$g_0\to -\infty$. 

 However, there is a strong peak in the variational scattering
intensity at some non-zero wavevector, which sharpens and increases in
height as the system is taken from the microemulsion phase into a
region of the phase diagram, where the lamellar phase is found in the
Monte Carlo simulations. Thus, in order to get an estimate for the
transition line, we use a Lindemann type criterion \cite{Lindemann} by
requiring that the dimensionless product $q\xi$ equal some fixed value
$w$ at the transition. We use here $w=5$; this choice is motivated by
the fact that values of $q\xi {> \atop \sim} 5$ have been observed
neither experimentally \cite{TeuStrey,ChenChang,Widom}, nor in the
simulations of the Ginzburg-Landau model \cite{GoKrausII}. The result
of this approximation is also shown in Fig. 5. Its qualitative
behavior is again in reasonable agreement with the Monte Carlo data.
The  location of the line of $q\xi=w$ obviously depends on the value
of $w$.

\section{Scattering Intensities}

 The scattering intensity in {\it bulk} contrast, $G_{\Phi\Phi}(k)
= <\Phi({\bf k}) \Phi(-{\bf k})>$, at different points in the phase
diagram is shown in Fig. 6. As we have already mentioned in the
previous section, the data of Fig. 6a have been used to determine the
cutoff parameter $\gamma_0$ of the variational approach. Note,
however, that $\gamma_0$ is the only parameter in the fit, which
determines the shape {\it and} the amplitude of the scattering intensity.
The same parameter $\gamma_0$ is then used to calculate the scattering
intensities at other points in the phase diagram, as shown in Fig. 6b
and 6c. The agreement with the Monte Carlo data is quite remarkable.

 The results of the self-consistent perturbation theory are shown
in Fig. 7, at the same points in the phase diagram as in Fig. 6. Given
the rather large calculational effort to obtain these curves, the
result is rather disappointing. The self-consistent perturbation theory
gives only a very weak peak at non-zero wavevector $k$. Although the
position of the peak is roughly correct, the ratio of peak height to
the scattering intensity at $k=0$ varies very little compared to the
Monte Carlo data.

 The Monte Carlo data for the {\it film} correlation function in real
space are shown in Fig. 8. We find that this correlation shows
{\it oscillations} as a function of distance $r$. This oscillatory behavior
has been observed explicitly so far only in lattice models for
microemulsions in one dimension
\cite{GoSch_latt1d,MatSul_latt1d,lengths_1d}. Oscillations are also
found in the Gaussian correlation function (\ref{film_posit}). This
correlation function is also shown in Fig. 8. Two main differences can
be recognized immediately: (i) the oscillations of the Monte Carlo
data are much more pronounced and (ii) the Monte Carlo data for
intermediate $r$ drop below the asymptotic value of $G_{film}$ for
$r\to\infty$, while the Gaussian correlation function does not (as
discussed above). These two differences have important consequences
when the film scattering intensity is calculated (by a numerical
Fourier transform of the data shown in Fig. 8). The results for this
scattering intensity are shown in Fig. 9a. Note that while the
variational correlation function decays monotonically, the Monte Carlo
result displays a small peak at some non-zero value of the wavevector
$k$. The latter behavior is just what is seen in experiment
\cite{SchuStrey,SMAP91,StreyMagid}. 

 The Fourier transform of Eq. (\ref{film_posit}) depends on a
single parameter, the dimensionless product $q\xi$. We can thus ask if
the Gaussian scattering intensity in film contrast shows a peak at
non-zero wavevector $k$ for {\it any} value of $q\xi$ (independent of
the value of $q\xi$ obtained by the variational method). The
correlation function (\ref{film_posit}) has its strongest oscillations
for large $q\xi$. However, even in the limit $q\xi\to\infty$, no peak
or shoulder appears at finite $k/q$. Thus, the Gaussian scattering
intensity in film contrast {\it never} has a peak at finite wavevector
$k$.

 An analytical expression for the film scattering intensity has
been obtained recently from a Ginzburg-Landau model with two scalar
order-parameter fields \cite{RCOBNB,sponge,GoSch94}, the concentration
difference between oil and water, $\Phi({\bf r})$, and the amphiphile
concentration, $\rho({\bf r})$. In this case, one finds for the
amphiphile-amphiphile correlation function \cite{GoSch94}
\begin{equation} \label{scatt_GS}
G_{\rho\rho}(k) = \frac{\chi_{\rho}}{1+(k\xi_{\rho})^2} +
 \left(\frac{\chi_{\rho}}{1+(k\xi_{\rho})^2}\right)^2 \ 
                                     \Gamma(k\xi, q\xi) 
\end{equation}
where 
\begin{eqnarray} \label{Gamma}
\chi_\rho \ \Gamma(x,y)
       &=& (\gamma_1 - \gamma_2 x^2)^2 \Lambda_{-}(x,y)  \\
       &+& 2\gamma_3 (\gamma_1 - \gamma_2 x^2)
       \left[(1 - y^2)
                    \Lambda_{-}(x,y) - y \Xi(x,y) \right] \nonumber \\
       &+& \gamma_3^2 \bigl[
                2 y^2 \left(\Lambda_{+}(x,y)-\Lambda_{-}(x,y)\right)
                        + (1-y^2)^2 \Lambda_{-}(x,y) \nonumber \\
       & & {\hskip 3 true cm}            
               - 2 y (1-y^2) \Xi(x,y) + y^2 \bigr],  \nonumber
\end{eqnarray}
with
\begin{eqnarray} \label{functdef1}
\Lambda_{\pm}(x, y) &=& \frac{1}{4 x}
  \left[ 2\arctan\left(\frac{x}{2}\right)
          \pm \arctan\left(\frac{4 x}{4+4y^2-x^2}\right)
                               \pm n\pi     \right] \\
\label{functdef2}
\Xi(x, y) &=& \frac{1}{4 x}
     \ln\left(\frac{4+(x+2y)^2}{4+(x-2y)^2}\right) \ .
\end{eqnarray}
Here, $n=0$ for $0\le x^2 < 4+4y^2$ and $n=1$ for $x^2 > 4+4y^2$. The
disadvantage of this two-order-parameter model is that it contains the
parameters $\gamma_1$, $\gamma_2$, $\gamma_3$, and $\xi_\rho$ in
addition to those which appear in our model (\ref{GL}), (\ref{f}),
(\ref{g}). We thus fit the analytic form (\ref{scatt_GS}) to our Monte
Carlo data, in order to compare the general form of the two results.
Here, the value of $q\xi$ is determined from a fit of the Monte Carlo
order-parameter correlation function to the Teubner-Strey form
(\ref{corr_var}). The two curves are shown in Fig. 9b. The agreement
is found to be very good for $0<k/q<3/2$. This implies in particular
that for $1 \ll k\xi \ll q\xi$ the Monte Carlo data show a $1/k$
behavior \cite{RCOBNB,sponge,GoSch94}. For larger values of $k$, the
expression (\ref{scatt_GS}) yields a more rapid decay than the Monte
Carlo data. The same discrepancy occurs when (\ref{scatt_GS}) is
compared with the experimental data, see Ref. \cite{GoSch94}. Thus,
our Monte Carlo results should describe the experimental behavior
even better than the perturbation theory of the
two-order-parameter model.

\section{Structure and Topology}

 A typical configuration of the $\Phi({\bf r})=0$ surfaces in a
microemulsion near the four-phase point is shown in Fig. 10. Note that
fluctuations of the interfaces on length scales {\it smaller} than the
typical domain size are clearly visible. We want to calculate now the
area (per unit volume), $S/V$, the mean curvature squared, $<H^2>$,
and the Euler-characteristic (per unit area), $\chi_E/S$ \cite{Euler}.
The results of the Monte Carlo simulations are shown in Figs. 11a, 12a
and 13a, and of the variational approach in Figs. 11b, 12b and 13b. It
can be seen that in all three cases the qualitative behavior of the
Monte Carlo data is reproduced by the variational results. The best
quantitative agreement is found for the Euler characteristic. There is
still a reasonable quantitative agreement for the area, while the mean
curvature squared is off by about a factor three (which is about the
same factor as for Gaussian random fields, see Section 4). 

 We have shown in Section 4 that the ratio $\chi_E V^2/S^3$ is
less sensitive to cutoff effects than the quantities $\chi_E/S$ and
$S/V$ themselves. Furthermore, it has been shown in Ref.
\cite{GoKrausII} by a simple scaling argument, which is valid beyond
Gaussian random fields, that this quantity characterizes the topology
of a microemulsion phase {\it independent} of the domain sizes of the
coherent oil- and water-regions. From the Monte Carlo simulations,
this ratio is found to be \cite{GoKrausII}
\begin{equation}
\frac{\chi_E V^2}{S^3} = - 0.176 \ \ \ \ \ \ \ \ \ \ 
                            {\rm (Monte\ Carlo)}
\end{equation}
while for Gaussian random fields, see Eq. (\ref{scale_KHS}),
\begin{equation}
\chi_E V^2/S^3= - \pi/16 = -0.196  \ \ \ \ 
                            {\rm (Gaussian\ random\ fields)}\ .
\end{equation}
Given the fact that the typical configurations of the order parameter
field look quite different, the similarity of these two values is
quite surprising.

\section{Discussion}

 We have studied in this paper the thermal fluctuations of
interfaces in microemulsions and sponge phases with the use of a
simple Ginzburg-Landau model for ternary amphiphilic systems. We have
employed Monte Carlo simulations, the variational method, and
self-consistent perturbation theory to calculate the scattering
intensities and correlation functions both in bulk and in film
contrast, as well as the area $S$, the mean curvature squared $<H^2>$,
and the Euler characteristic $\chi_E$ of the internal interfaces.

 The self-consistent perturbation theory is unable to produce a
strong peak in the scattering intensity with bulk contrast at non-zero
wavevector; it is thus limited in its application to weakly structured
microemulsions. A comparison of the Monte Carlo method and the
variational approach shows very good agreement for the scattering
intensity in bulk contrast, for the phase transition between the
microemulsion and the oil-rich and water-rich phases, and for the
dimensionless ratio $\chi_E V^2/S^3$. This is a very useful result,
since it indicates that in order to calculate phase diagrams of the
Ginzburg-Landau model for other sets of parameters, time-consuming
Monte Carlo simulations in search for phase transitions can be
avoided. However, the variational approach fails for the scattering
intensity in film contrast, and for the phase transition between
microemulsion and lamellar phase. In particular, the variational
approach is unable to produce a peak of the scattering intensity in
film contrast at finite wavevector $k$. This is a serious deficiency
of the Gaussian model. It shows that {\it level surfaces of Gaussian
random fields do not accurately describe the structure of bicontinuous
microemulsions}. The Monte Carlo results of our Ginzburg-Landau model,
on the other hand, are in very good agreement with experimental
results.

\bigskip  
\noindent 
{\bf Acknowledgements:} 
Stimulating interactions with M. Kraus are greatfully acknowledged.
We also thank R. Hausmann, M. Schick and H. Wagner for helpful
discussions. This work was supported in part by the Deutsche
Forschungsgemeinschaft through Sonderforschungsbereich 266, and by
Cray Research through the University Research \& Development Grant
Program.

\newpage

\vskip .5cm
{\Large{\bf Figure Captions}}

\vskip 0.5cm

Figure 1: Auto-correlation function, ${\tilde G}(t)$, as a function of
time $t$ (in units of Monte Carlo steps per site), for the parameters
$g_0=-2$ and $f_0=0.5$ ($N=27$, $a_0=0.8 \ell_0$).

\vskip 0.5cm

Figure 2: (a) Dyson equation for the self-energy $\Sigma$. (b)
Two-loop expansion of the vertex function $\Gamma^{(4)}$. (c)
Two-loop expansion of the vertex function $\Gamma^{(6)}$. In these
Feynman diagrams, the full lines indicate the full propagator 
$G({\bf k})$; vertices without a circle are the bare vertex functions
$\Gamma_{bare}^{(4)}$ and $\Gamma_{bare}^{(6)}$, vertices with a
full circle denote the renormalized vertex functions.

\vskip 0.5cm

Figure 3: (a) Area (per unit volume), $S/V$, (b) average mean
curvature, $<H>$, (c) average mean curvature squared, $<H^2>$, and (d)
Euler characteristic (per unit area), $\chi_E/S$, of level surfaces of
Gaussian random fields. Note the logarithmic scale of the ordinate in
(a). The dashed lines are linear fits of the Monte Carlo data ($N=27$,
$a_0=0.8 \ell_0$). The full lines are the exact results of the
continuum model with cutoff parameter $\gamma_0=2.5$.

\vskip 0.5cm

Figure 4: The scaled Euler characteristic, $- \chi_E V^2/S^{3}$, as a
function of the scaled mean curvature, $ - <H>^2 S/(2\pi\chi_E)$, of
level surfaces of Gaussian random fields. The diamonds are the Monte
Carlo data, the full line is the exact result (\ref{scale_fun}). For a
better comparison of the shape of the two curves, we also show the
exact result multiplied by a factor $1.053$ (dashed line).

\vskip 0.5cm

Figure 5: Phase diagram of the Ginzburg-Landau model (\ref{GL}). Full
lines give the mean-field results, dashed lines the Monte Carlo
results \cite{GoKrausII}. In the variational approach, the transition
from oil-water coexistence to the microemulsion is indicated by one
dotted line (narrow spacing), the spinodal (where the microemulsion
looses its metastability) by another dotted line (wide spacing).
Finally, the variational estimate of the transition  microemulsion -
lamellar is the dashed-dotted line.

\vskip 0.5cm

Figure 6: Scattering intensity in bulk contrast. The full lines are
the results of the variational approach, the diamonds are the Monte
Carlo data ($N=45$, $a_0=0.6 \ell_0$). (a) $g_0=-1.0$, $f_0=0.0$.
The data at this point in the phase diagram are used to determine the
cutoff parameter $\gamma_0=1.5$. (b) $g_0=-2.0$, $f_0=0.5$. (c)
$g_0=-2.5$, $f_0=0.675$.

\vskip 0.5cm

Figure 7: Scattering intensity in bulk contrast as obtained from the
self-consistent screening approximation. The parameters are
$g_0=-1.0$, $f_0=0.0$ (full line), $g_0=-2.0$, $f_0=0.5$ (dashed
line), and $g_0=-2.5$, $f_0=0.675$ (dotted line).

\vskip 0.5cm

Figure 8: Film correlation function $G_{film}(r)$ in real space, for
$g_0=-2.5$ and $f_0=0.675$, with film thickness $\epsilon=0.1 \Phi_b$.
The full line is the variational result (with $\gamma_0=1.5$), the
Monte Carlo data ($N=27$, $a_0=0.8 \ell_0$) are given by diamonds.

\vskip 0.5cm

Figure 9: Scattering intensity in film contrast. The parameters are
the same as in Fig. 8. (a) A comparison of Monte Carlo simulation
(full line) and the variational approach (dashed line). (b) A
comparison of the Monte Carlo data (full line) with the results of the
two-order-parameter Landau model, Eq. (\ref{scatt_GS}), with
$\gamma_1=36.1$, $\gamma_2=-0.21$, $\gamma_3=0.53$, and
$\xi_\rho/\xi=0.072$ (dashed line). The value of $q\xi=9.71$ is
obtained by fitting the Monte Carlo results for scattering intensity
in bulk contrast to the Teubner-Strey form (\ref{corr_var}).

\vskip 0.5cm

Figure 10: Typical configuration of the $\Phi({\bf r})=0$ surfaces in
a microemulsion. The parameters are $g_0=-2.0$, $f_0=0.75$, $a_0=0.6
\ell_0$, and $N=45$. The figure shows only a part (of size $36\times
36 \times 36$) of the total lattice. The two sides of the interface
are colored differently, dark on the oil-rich side, light on the
water-rich side.

\vskip 0.5cm

Figure 11: Contour plot of the area $S$ of internal oil/water
interfaces ($\Phi({\bf r})=0$) in the microemulsion or sponge phase,
as obtained from (a) Monte Carlo simulations ($N=27$, $a_0=0.8
\ell_0$), and (b) the variational method. The bold lines indicate the
positions of phase transitions and spinodals, compare Fig. 5. 

\vskip 0.5cm

Figure 12: Contour plot of the mean curvature squared $<H^2>$ of
internal oil/water interfaces ($\Phi({\bf r})=0$) in the microemulsion
or sponge phase, as obtained from (a) Monte Carlo simulations ($N=27$,
$a_0=0.8 \ell_0$), and (b) the variational method. The bold lines
indicate the positions of phase transitions and spinodals, compare
Fig. 5.

\vskip 0.5cm

Figure 13: Contour plot of the Euler characteristic $\chi_E$ of
internal oil/water interfaces ($\Phi({\bf r})=0$) in the microemulsion
or sponge phase, as obtained from (a) Monte Carlo simulations ($N=27$,
$a_0=0.8 \ell_0$), and (b) the variational method. The bold lines
indicate the positions of phase transitions and spinodals, compare
Fig. 5.
\newpage
\pagestyle{empty}
\vspace*{-4cm} \hspace*{-3cm} \epsfbox{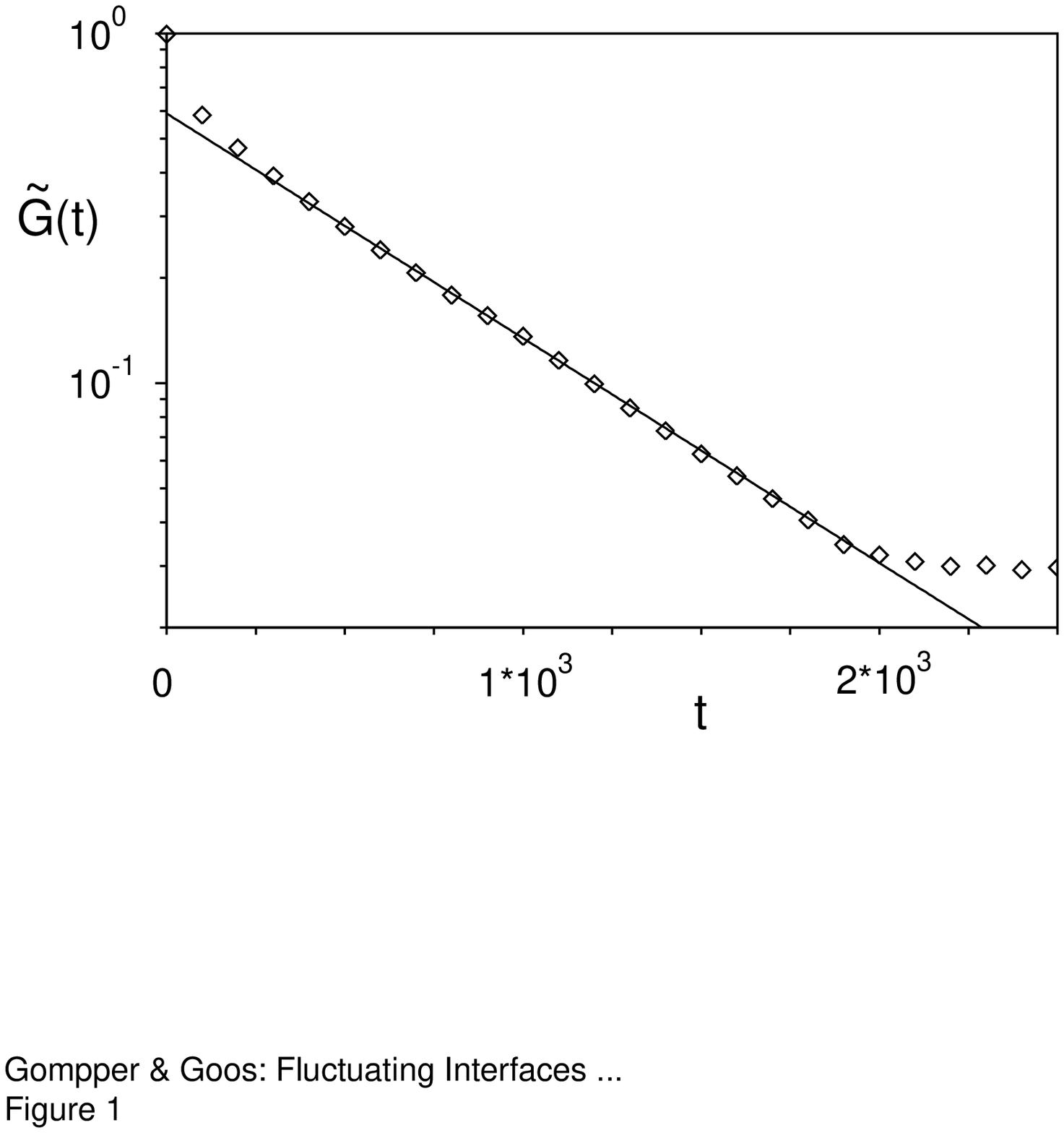}
\newpage
\vspace*{-3.5cm} \hspace*{-3cm} \epsfbox{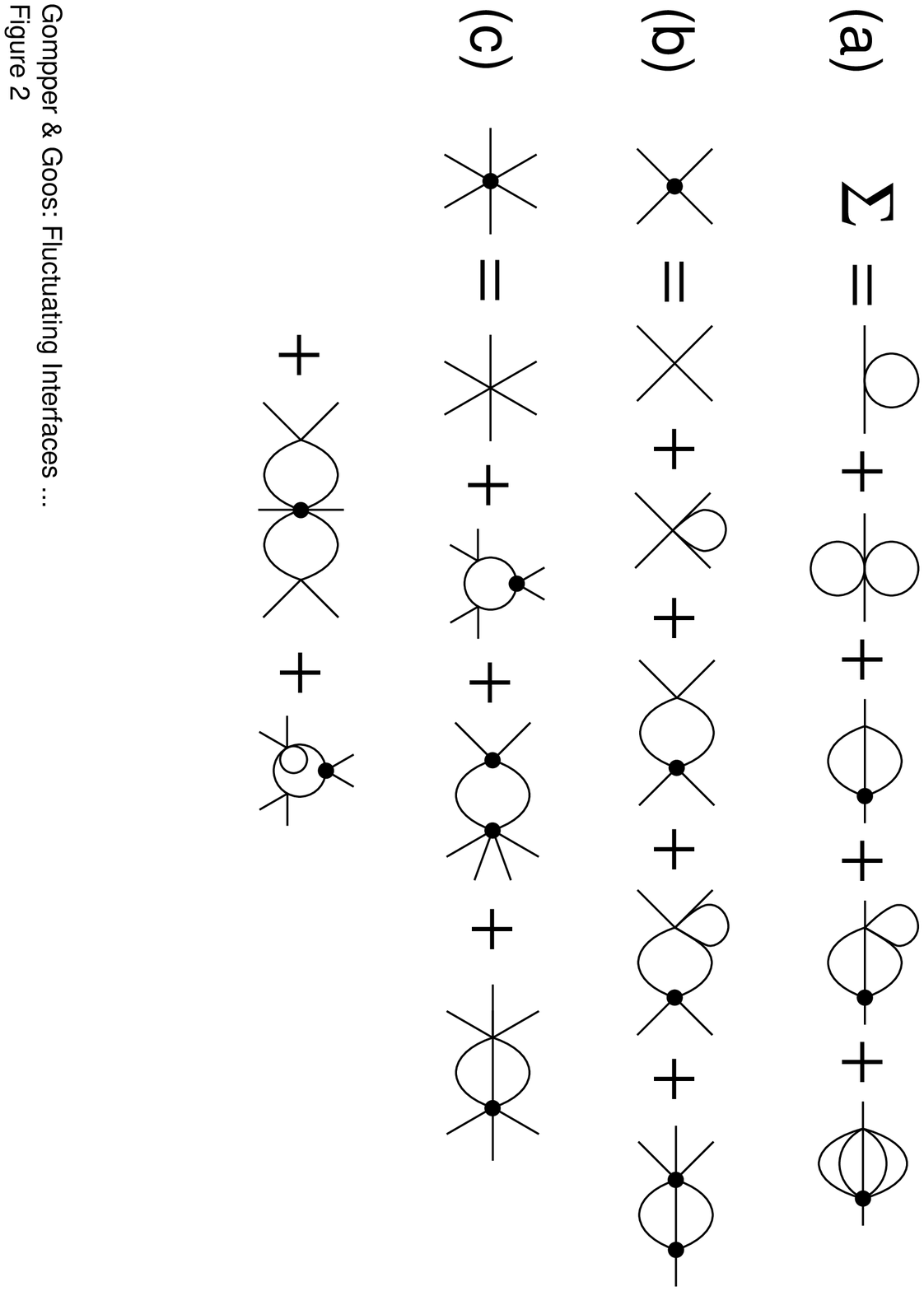}
\newpage
\vspace*{-4cm} \hspace*{-3cm} \epsfbox{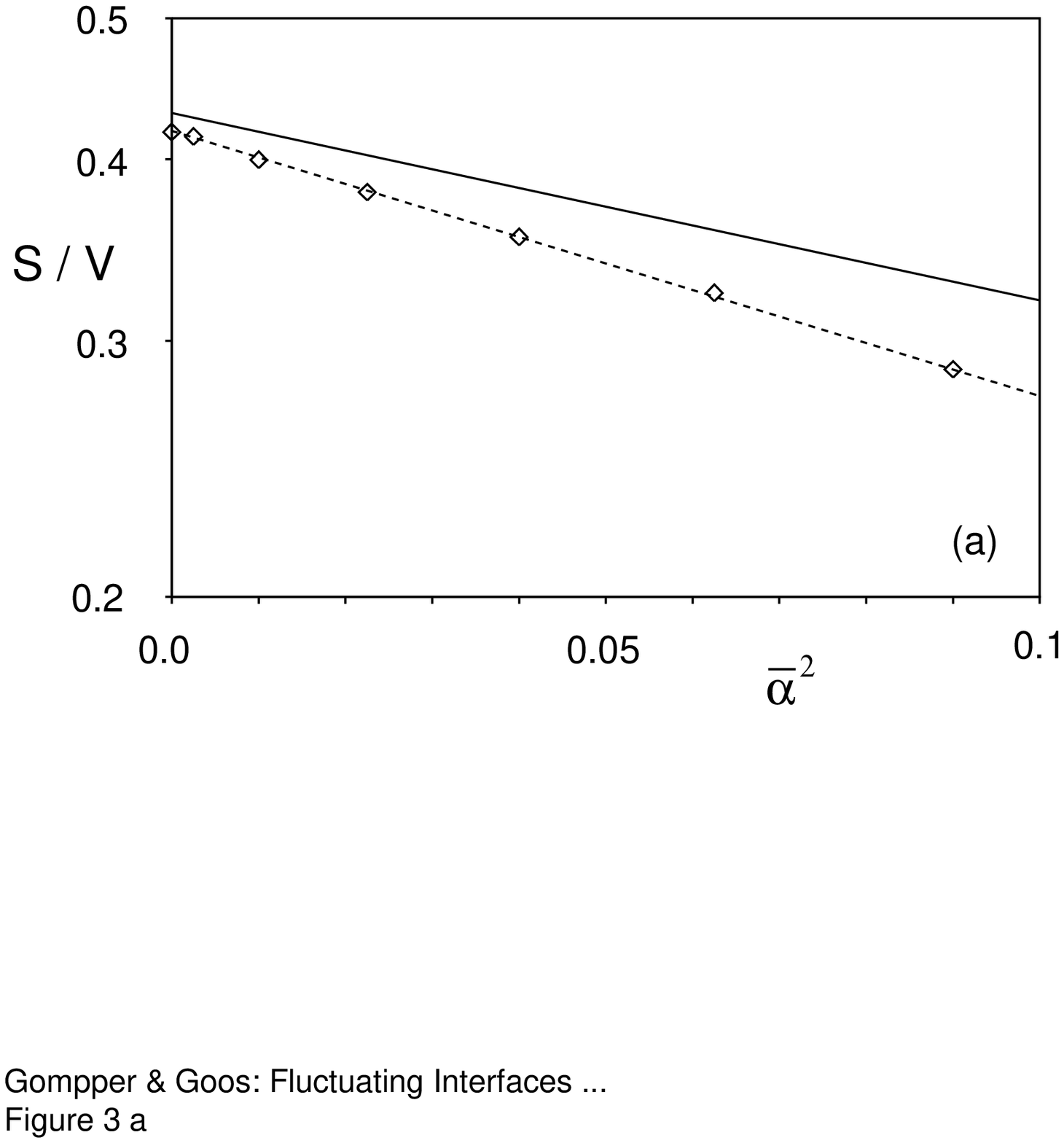}
\newpage
\vspace*{-4cm} \hspace*{-3cm} \epsfbox{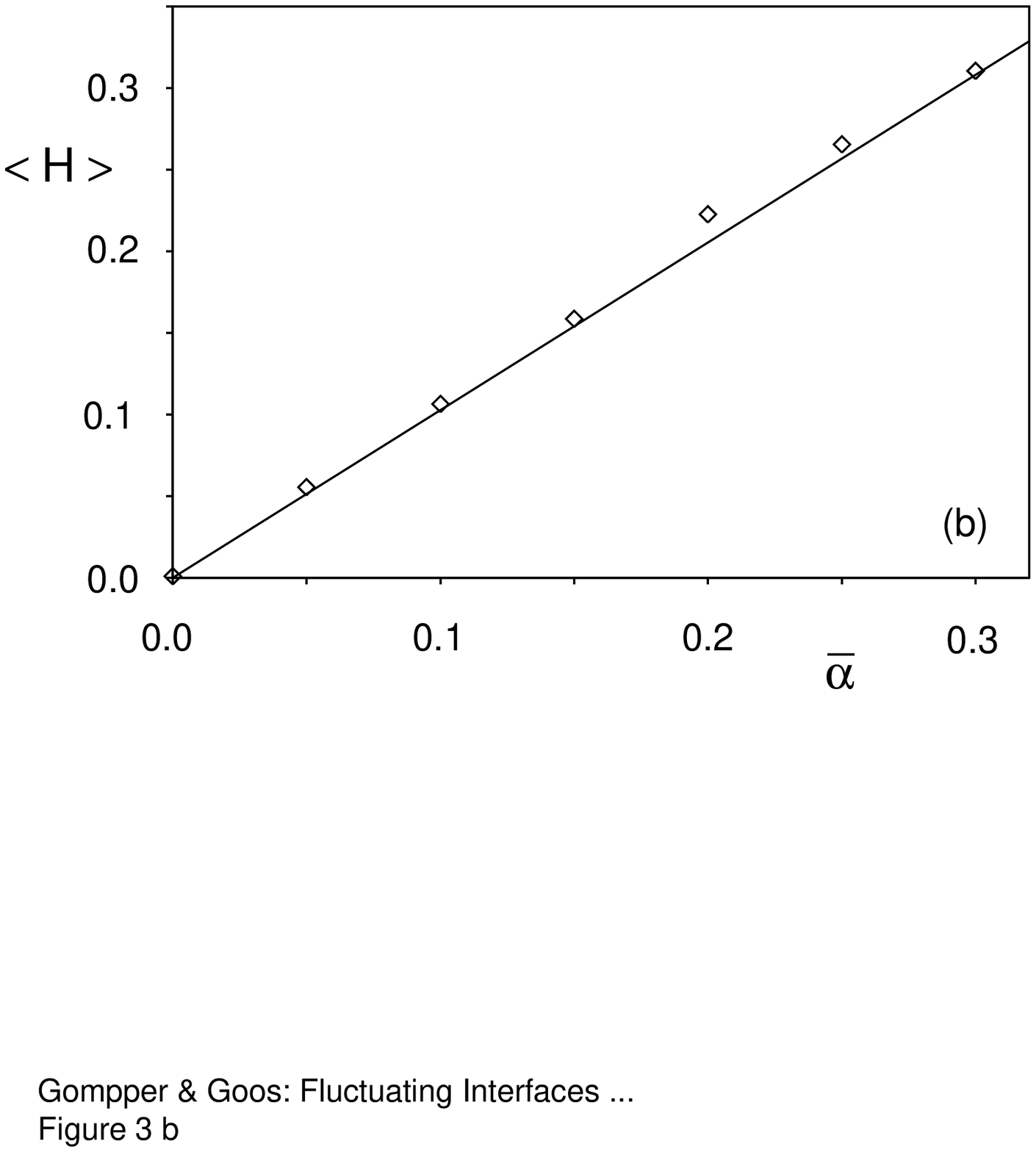}
\newpage
\vspace*{-4cm} \hspace*{-3cm} \epsfbox{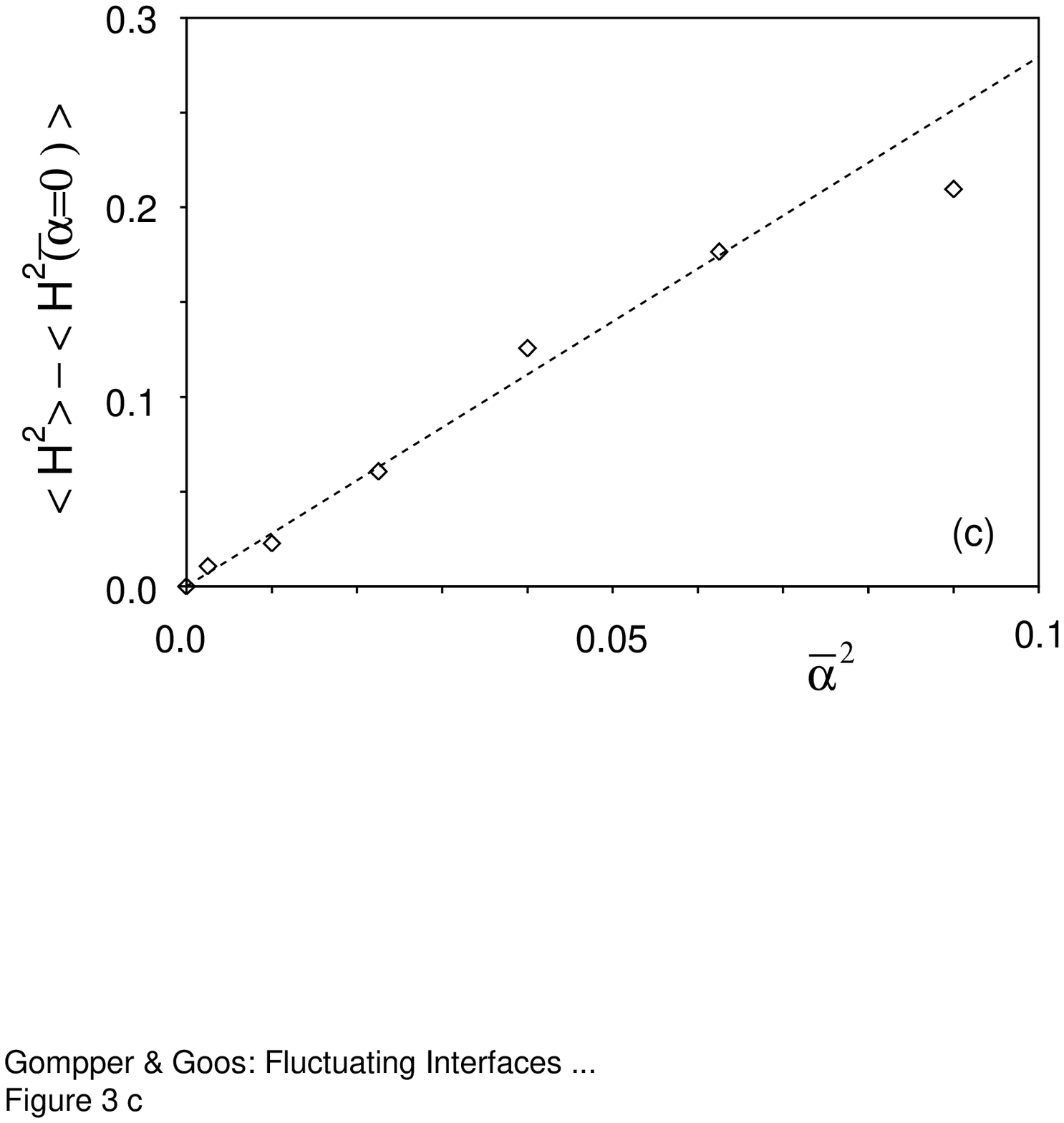}
\newpage
\vspace*{-4cm} \hspace*{-3cm} \epsfbox{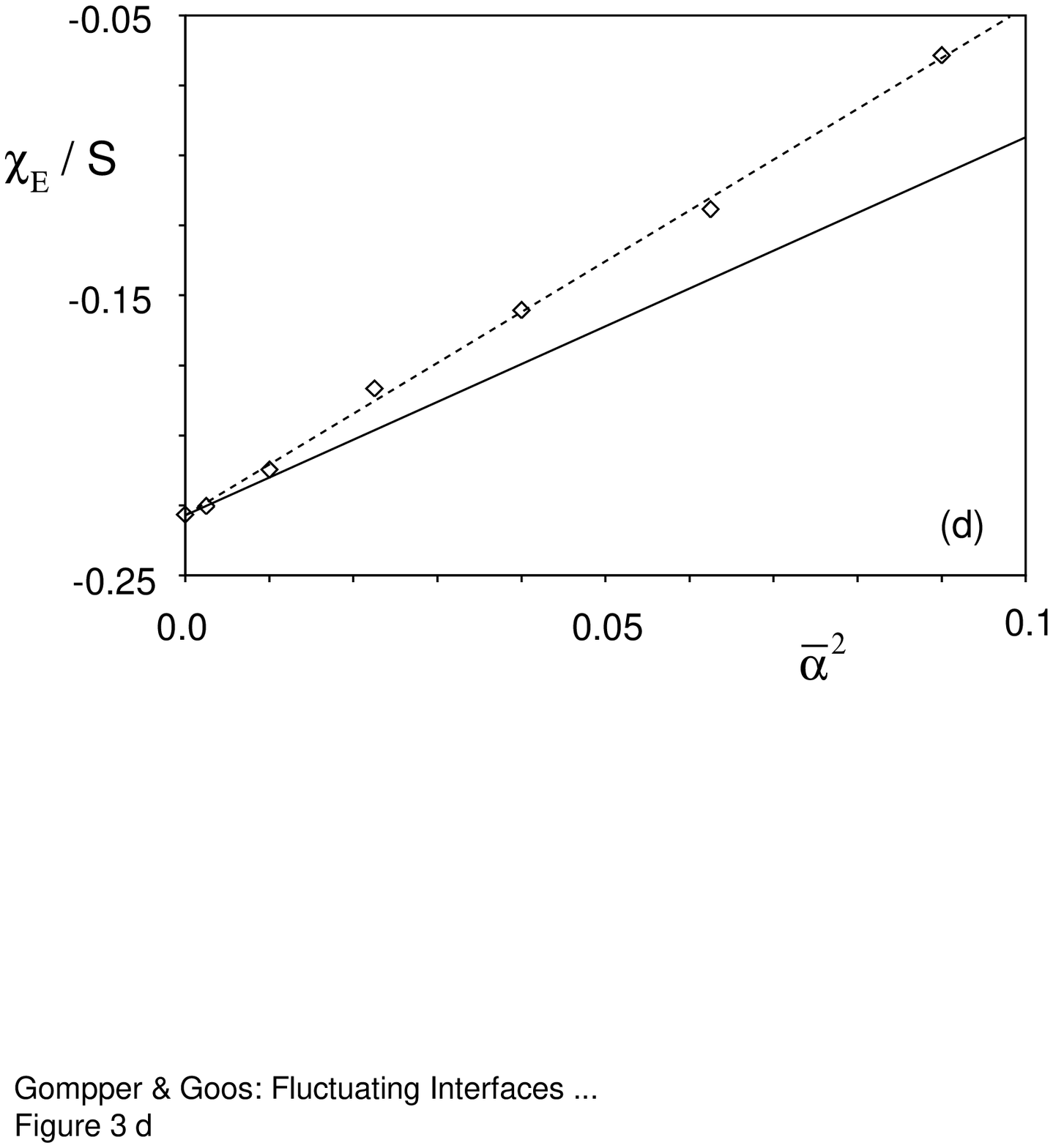}
\newpage
\vspace*{-4cm} \hspace*{-3cm} \epsfbox{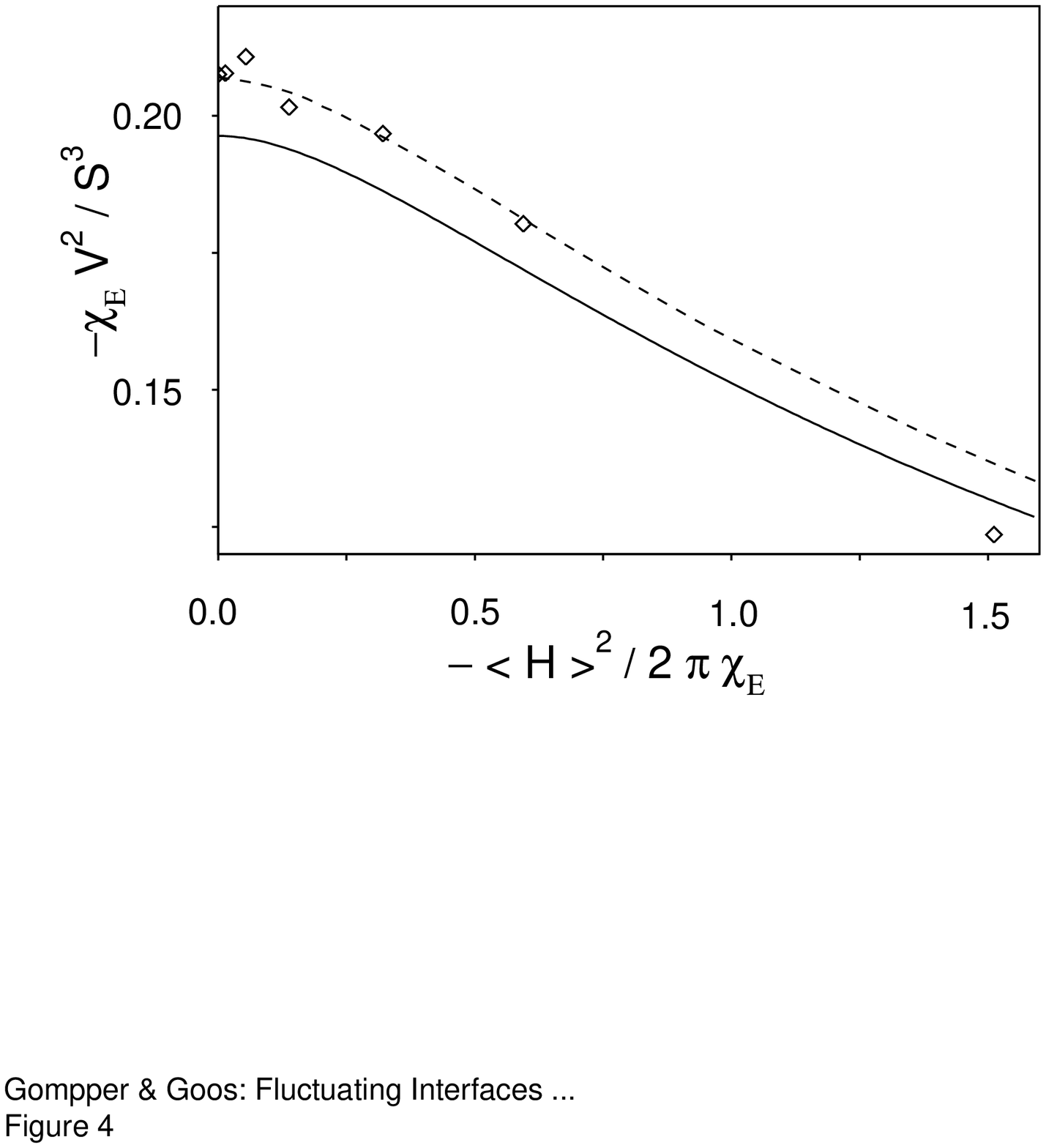}
\newpage
\vspace*{-4cm} \hspace*{-3cm} \epsfbox{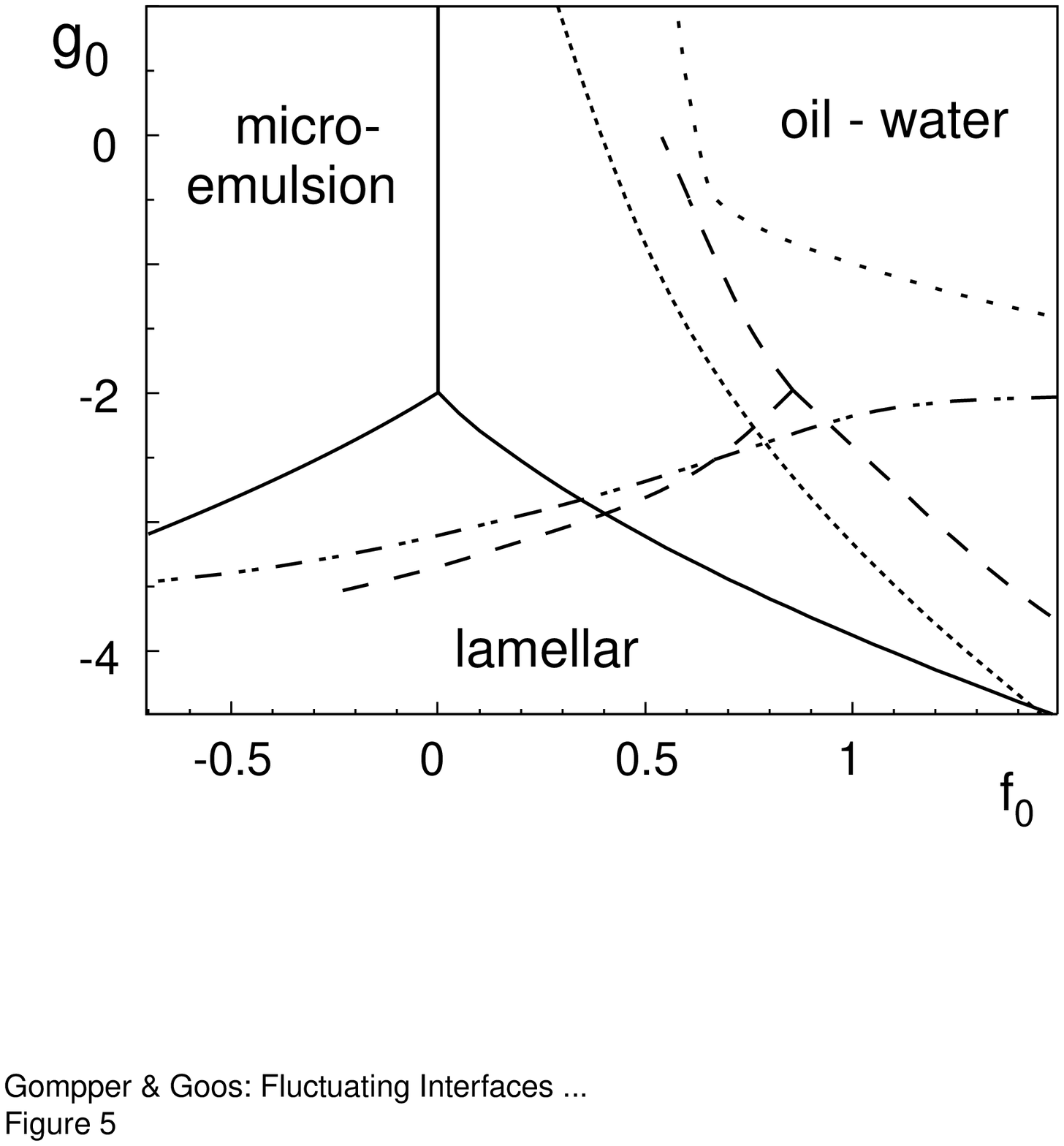}
\newpage
\vspace*{-4cm} \hspace*{-3cm} \epsfbox{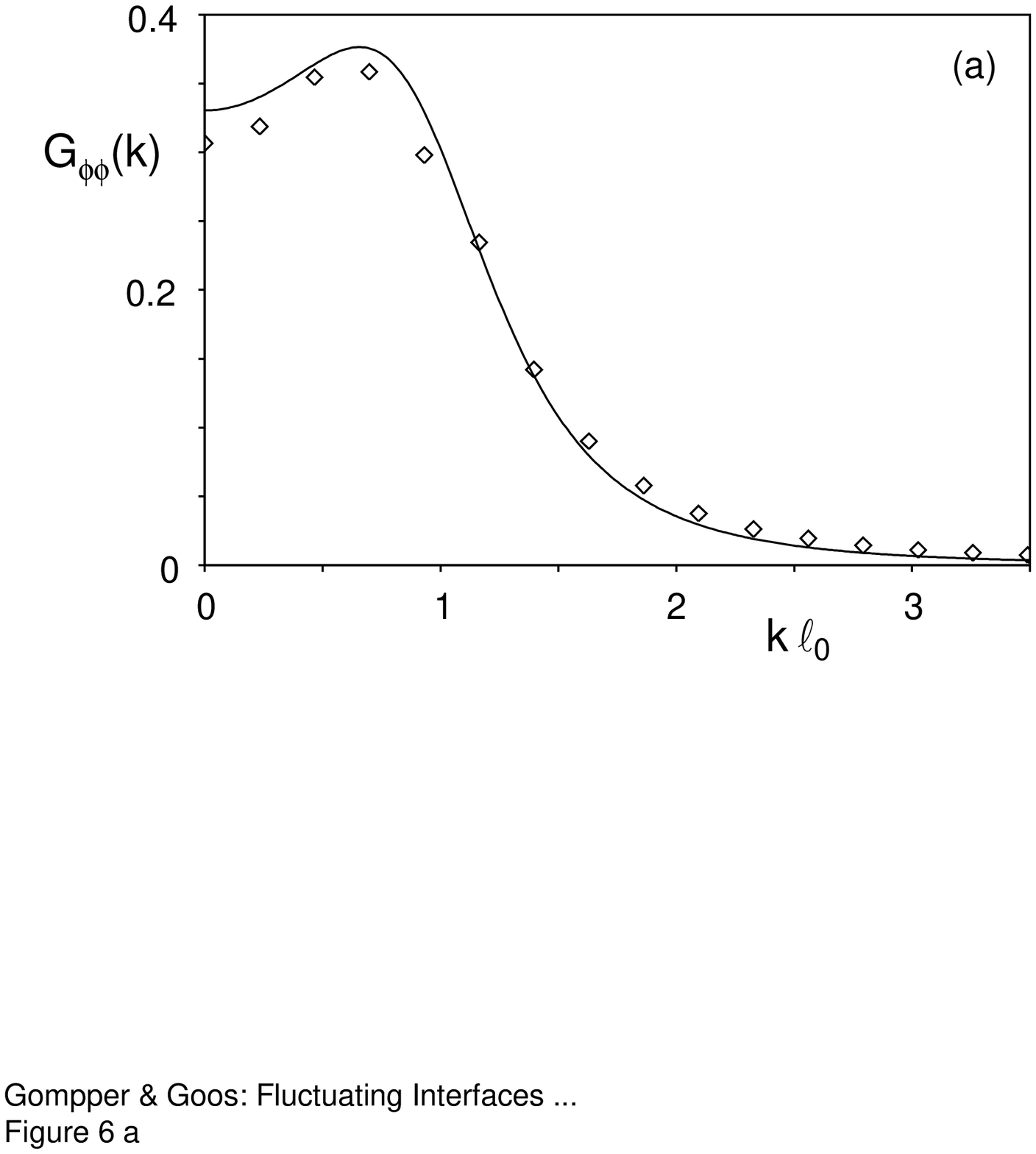}
\newpage
\vspace*{-4cm} \hspace*{-3cm} \epsfbox{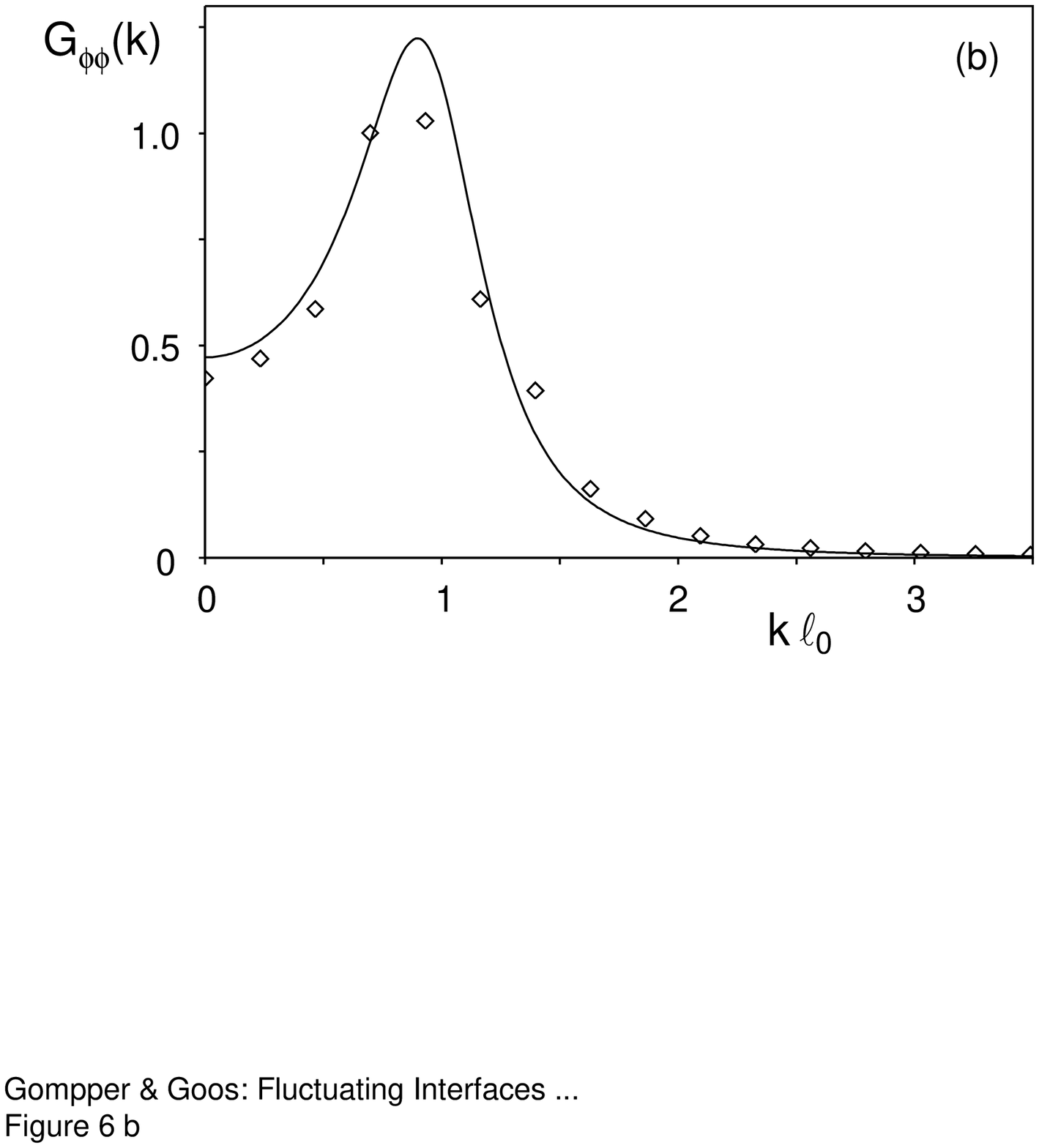}
\newpage
\vspace*{-4cm} \hspace*{-3cm} \epsfbox{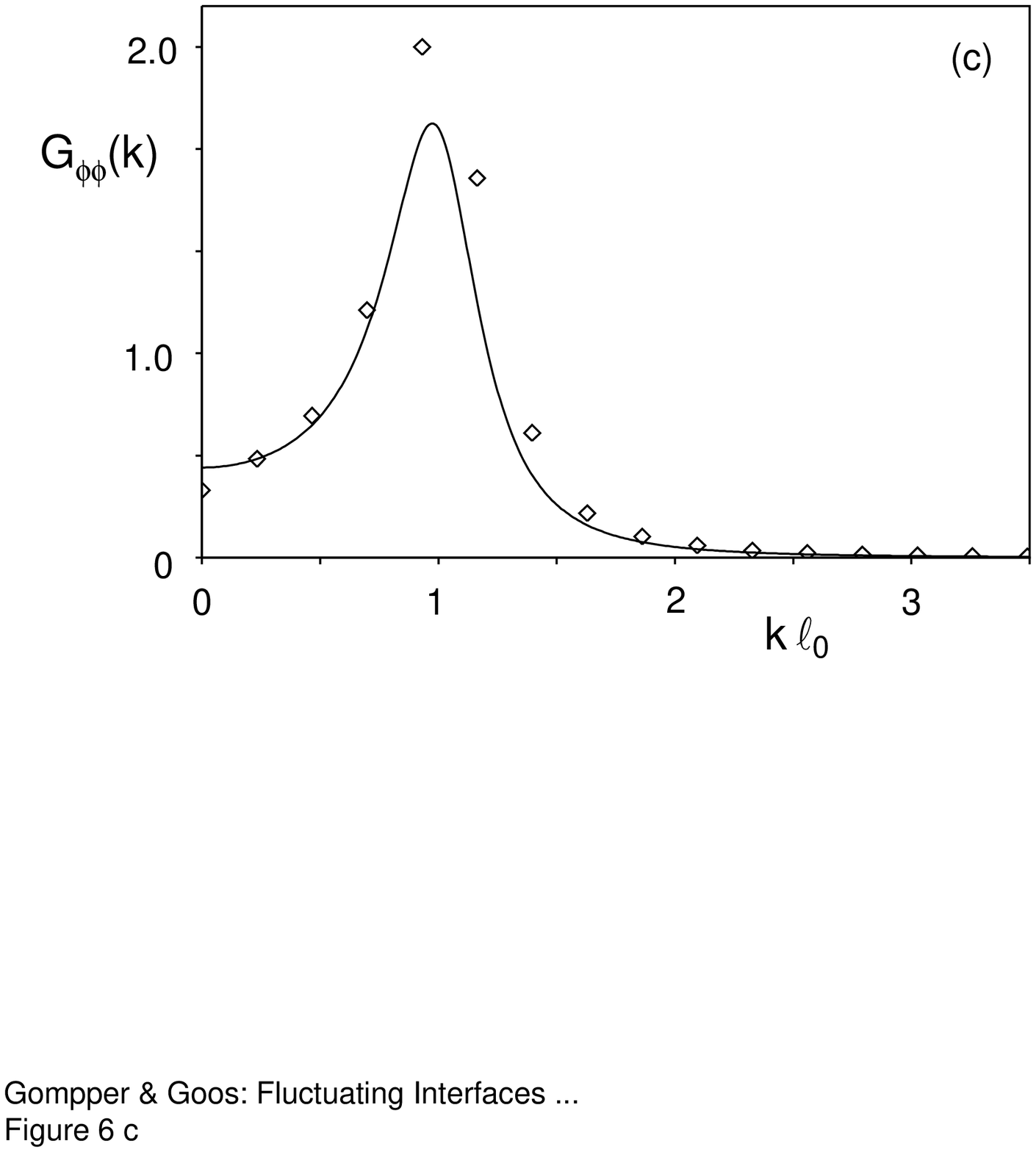}
\newpage
\vspace*{-4cm} \hspace*{-3cm} \epsfbox{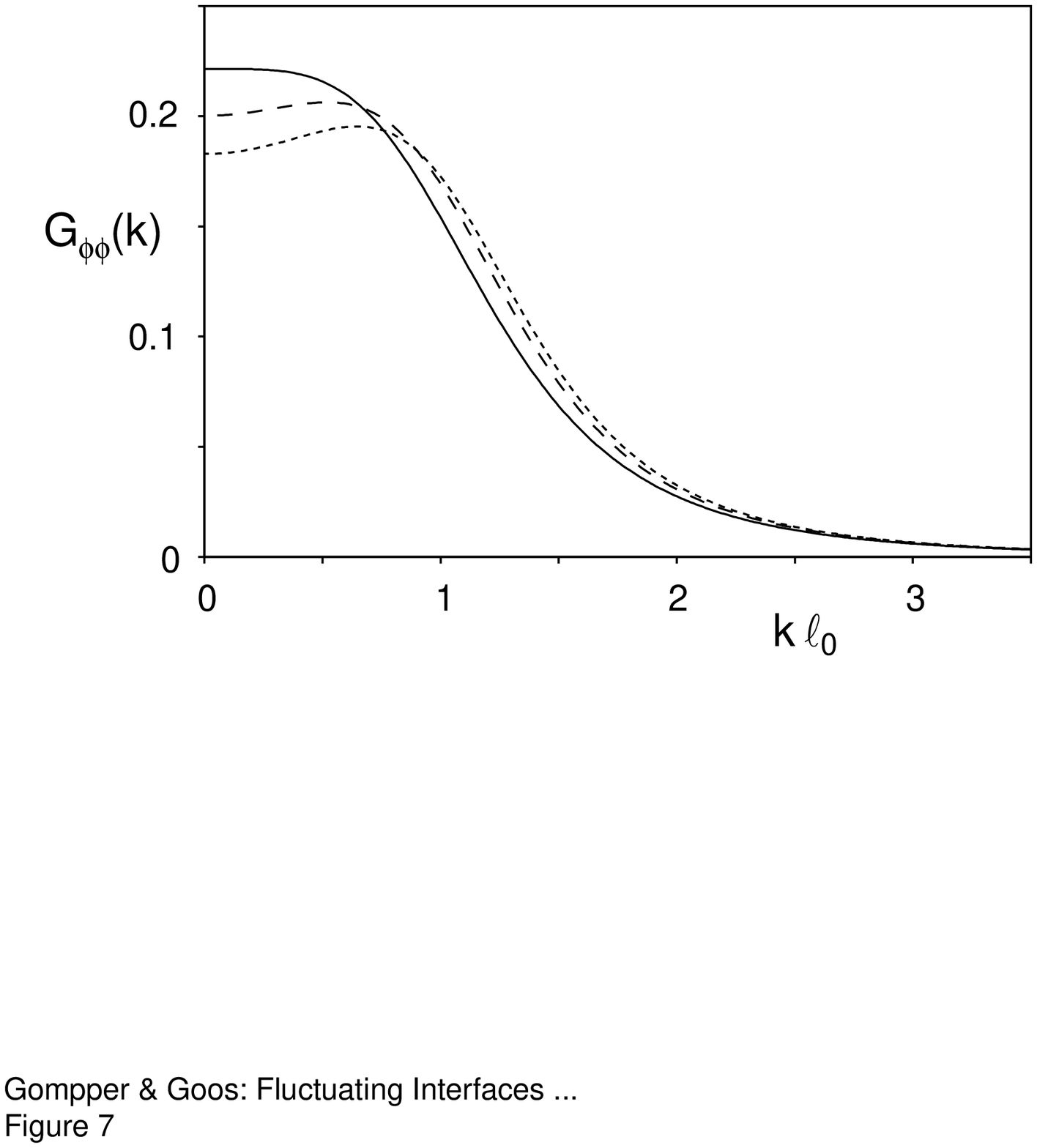}
\newpage
\vspace*{-4cm} \hspace*{-3cm} \epsfbox{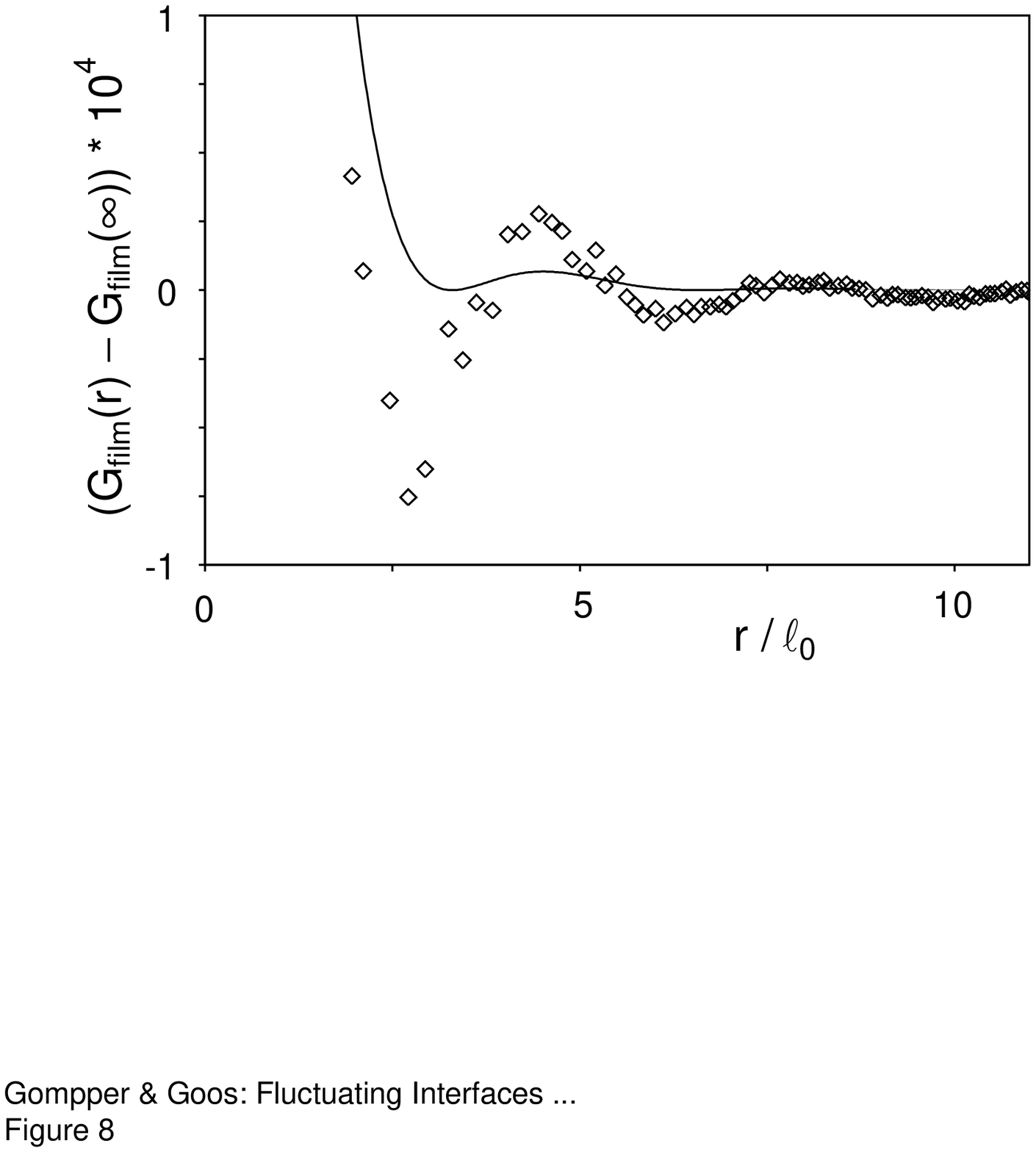}
\newpage
\vspace*{-4cm} \hspace*{-3cm} \epsfbox{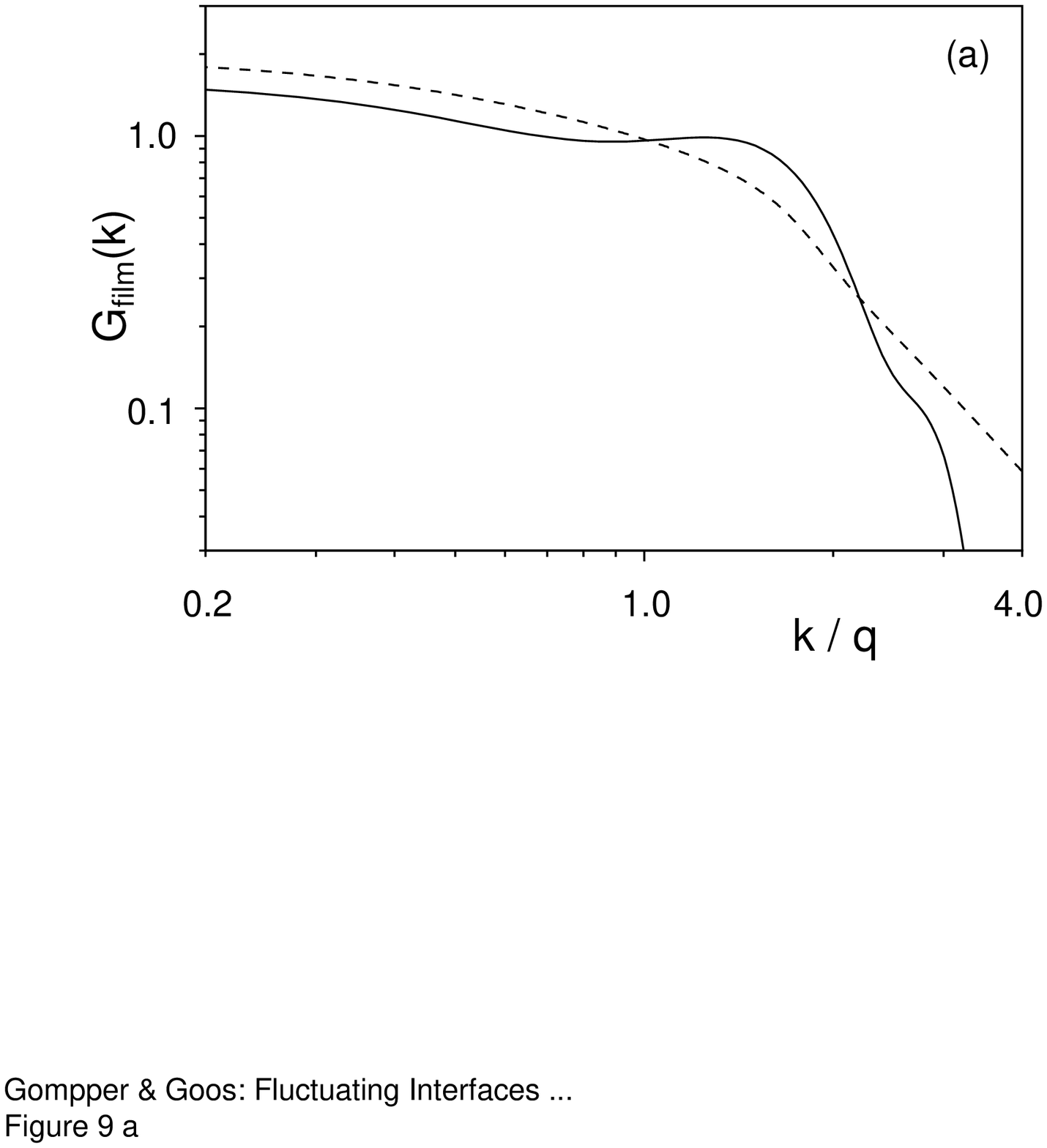}
\newpage 
\vspace*{-4cm} \hspace*{-3cm} \epsfbox{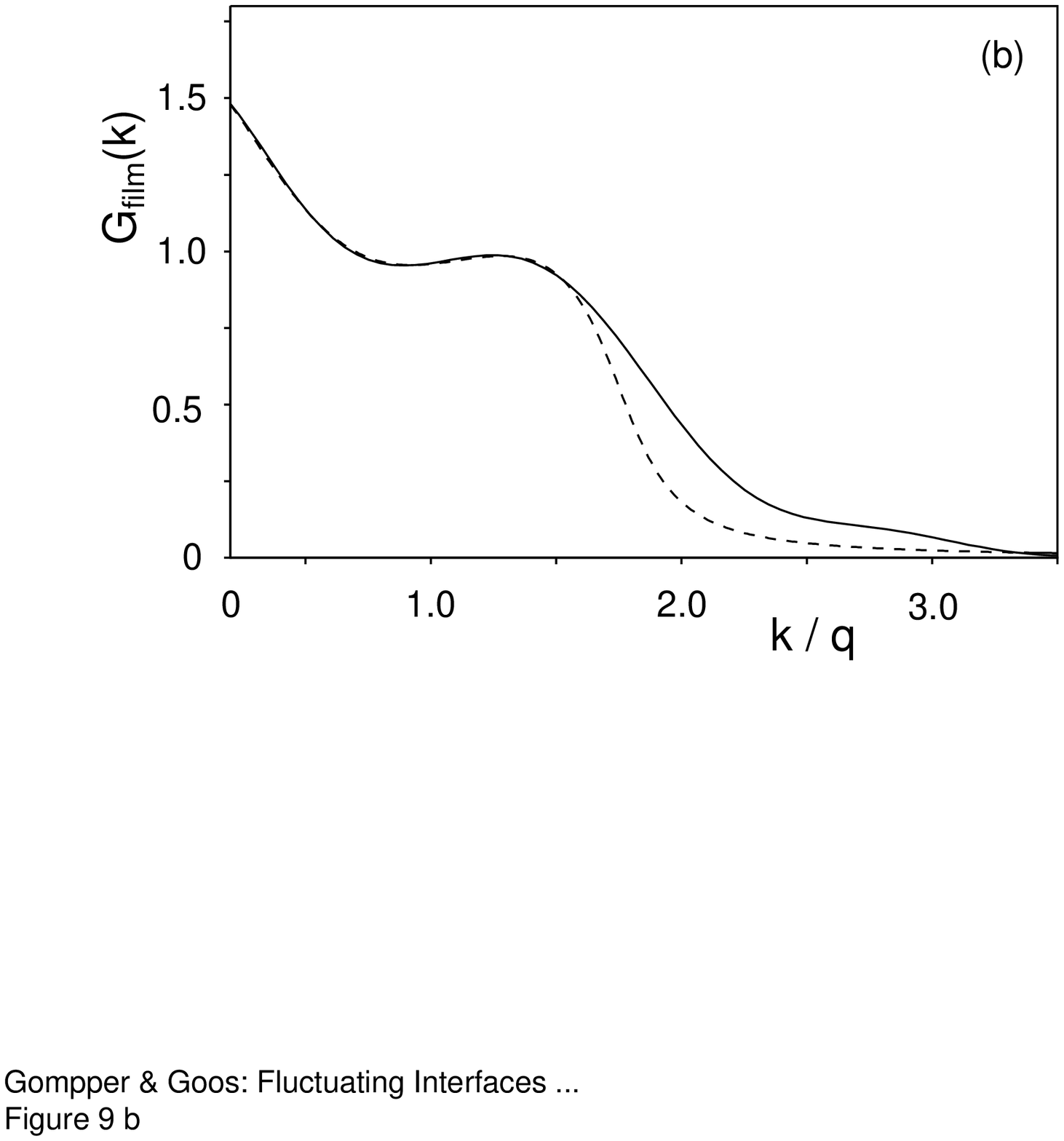}
\newpage
\vspace*{-4cm} \hspace*{-3.3cm} \epsfbox{fig10.ps}
\newpage
\vspace*{-4cm} \hspace*{-3cm} \epsfbox{fig11.ps}
\newpage
\vspace*{-4cm} \hspace*{-3cm} \epsfbox{fig12.ps}
\newpage
\vspace*{-4cm} \hspace*{-3cm} \epsfbox{fig13.ps}


\begin{thebibliography}{99}
\bibitem{Euler} There are several possibilities to define the Euler
characteristic on the lattice, depending on how the surfaces which
touch at edges or corners are treated \cite{Mecke}. We use here a
symmetrized version, where in these cases the surfaces are counted as 
connected and disconnected with probability $1/2$. Another possibility
is discussed in Ref. \cite{HofKlein}.
\bibitem{general_barphi} The generalization of these results to the
case $<\Phi> = {\overline \Phi}$ is straightforward.
\end{thebibliography}
\end{document}